\documentstyle[11pt,aaspp]{article}

\begin{document}
{\it Astrophysical Journal Supplement Series, in press}
\title{Atomic data for astrophysics. I.
Radiative recombination rates for H-like, He-like,
Li-like and Na-like ions over a broad range of temperature}

\author{D. A. Verner and G. J. Ferland}
\affil{Department of Physics and Astronomy, University of Kentucky,
Lexington, KY 40506; verner@pa.uky.edu, gary@pa.uky.edu}

\begin{abstract}
We present new calculations and analytic fits to the rates of
radiative recombination towards H-like, He-like, Li-like and Na-like ions
of all elements from H through Zn ($Z=30$). The fits are valid over a wide
range of temperature, from 3~K to $10^9$~K.
\end{abstract}


\section{Introduction}
Astrophysical plasmas are generally dilute and far from thermodynamic
equilibrium.  As a result their physical state and resulting spectra
are determined by the balance set by a host of microphysical processes.
Total recombination rate coefficients are essential for the
prediction of the ionization balance, among other things.
Recombination at low densities proceeds through three
different physical processes: radiative recombination,
${\rm A}^{+n} + e^- \rightarrow {\rm A}^{+n-1} + h\nu$; dielectronic
recombination,
${\rm A}^{+n} + e^- \rightarrow {\rm A}^{*,+n-1} \rightarrow {\rm A}^{+n-1}
+ h\nu$; and charge transfer recombination, ${\rm A}^{+n} + {\rm H}^0
\rightarrow {\rm A}^{+n-1} + {\rm H}^+$.
This paper focuses on radiative recombination for the four isoelectronic
sequences where this process should be predominate at nebular
($T \lesssim 10^4$~K) temperatures. Referring to the radiative recombination
for an atom/ion, we mean the recombination which forms this species. E.g.,
``radiative recombination for \ion{He}{1}'' means the recombination from
\ion{He}{2} to \ion{He}{1}.

During the last two decades,
the power-law fits to the radiative recombination rates (Aldrovandi \&
Pequignot 1973, hereafter AP; Shull \& Van Steenberg 1982,
Arnaud \& Rothenflug 1985, Landini \& Monsignori Fossi 1990, 1991)
have been extensively used for
astrophysical applications. The AP calculations were made
for He, C, N, O, Ne, Mg, Si and S.
The total radiative
recombination coefficient was considered as a sum of recombination
coefficients to the ground level and to all the excited levels.
The ground level recombination coefficients were evaluated by use
of the Milne relation (see, e.g., Bates \& Dalgarno 1962) from the
photoionization cross sections known in early
seventies; for many species, extrapolations of the cross sections along
isoelectronic sequences were necessary due to lack of data.
For all the excited levels,
the hydrogenic approximation was assumed. The calculated
radiative recombination coefficients were fitted by
\begin{equation}
\alpha_{\rm r}(T)=A\left(\frac{T}{10^4~{\rm K}}\right )^{-\eta},
\end{equation}
where $T$ is temperature, $A$ and $\eta$ are the fitting
parameters.
Shull \& Van Steenberg (1982) calculated and fitted by formula (1)
the radiative recombination rates for all Fe ions. They also interpolated
the rate coefficients for Ar, Ca and Ni. Interpolations of the rates
for Al, Na, F, P, Cl, K, Ti, Cr, Mn and Co were done by
Landini \& Monsignori Fossi (1990, 1991).
Arnaud \& Rothenflug (1985) improved the AP and Shull \& Van Steenberg (1982)
rates for recombination
towards He-like, Li-like and Be-like ions, using the same power-law
fitting formula (1).

Gould (1978) introduced corrections to the hydrogenic formula for
the radiative recombination coefficients to the excited levels of nonhydrogenic
species. He provided more accurate total radiative recombination
coefficients for the first four ionization states of C, N, O, Ne,
Mg, Si, S, Ar at $T=10^4$~K, and for the neutral species of these
elements at $T=10^2$~K. Pequignot, Petitjean, \& Boisson (1991)
used updated photoionization cross sections from the ground and first
excited ($n=2$ and $n=3$) states of all ions of He, C, N, O, and Ne for
calculation of the radiative recombination coefficients. All levels
$n>3$ were assumed to be hydrogenic. The results were fitted by
the 4-parametric formula
\begin{equation}
\alpha_{\rm r}(T)=10^{-13}z\frac{at^b}{1+ct^d} \,\,\, {\rm cm}^3{\rm s}^{-1},
\end{equation}
where $z$ is the
ionic charge ($z=1$ for recombination towards neutral state),
$t=10^{-4}T/z^2$, and $a$, $b$, $c$, and $d$ are the fitting
parameters. These fits are
valid for the temperature range $40z^2 \le T \le 4\times10^4z^2$~K.
Arnaud \& Raymond (1992) calculated iron radiative recombination
rates of \ion{Fe}{15} through \ion{Fe}{26} using the photoionization
cross sections from the ground and all the excited states with
$n \le 5$ presented by Clark, Cowan, \& Bobrowicz (1986). The rates
coefficients were fitted by a modified form of Eq. (1);
\begin{equation}
\alpha_{\rm r}(T)=A\left(\frac{T}{10^4~{\rm K}}\right )^
{-\alpha-\beta\log_{10}(T/10^4~{\rm K})},
\end{equation}
where $A$, $\alpha$ and $\beta$ are the fitting parameters.

This paper is part of a series concerned with the basic data needed
to fully simulate spectral formation by all stages of ionization of
the first thirty elements.
Reliable rates are needed over the full temperature range
3 K $\le T \le 10^9$ K.
During the last years, new extensive calculations of the photoionization
cross sections have been performed and published. Verner \& Yakovlev
(1995) presented a complete set of analytic fits to the partial
Hartree-Dirac-Slater (HDS) photoionization cross sections for the ground
state shells of all atoms and ions of elements from H to Zn.
These fits are very accurate far from the ionization thresholds of
the outer shells and at higher energies above the subsequent inner
shell ionization edges. However, the HDS method may give significant
errors near the thresholds of the outer shells, especially for neutral
and low-ionized species. Use of the fits to the HDS photoionization
cross sections for evaluation of the radiative recombination rates
may lead to large errors in the low-temperature rates for neutral
and low-ionized species. The recent Opacity Project (OP) calculations
(Seaton et al. 1992) give more accurate low-energy photoionization
cross sections obtained by the R-matrix method (Burke et al. 1971)
for all atoms and ions of elements with $Z \le 14$ and $Z=16$, 18, 20, 26
($Z$ is atomic number).
Unfortunately the OP calculations do not include high-energy
photoionization cross sections;  these are required for high-temperature
radiative recombination rates. Moreover, the OP database TOPbase
(Cunto et al. 1993) contains only total photoionization cross sections
which are not appropriate for use with the Milne relation.

For the most ionic species, the total recombination rates are
dominated by dielectronic recombination
(Burgess 1964; Nussbaumer \& Storey
1984, 1986, 1987). The exceptions are H-like ions which are not
subject to dielectronic recombination at all, and
He-like, Li-like
and Na-like species which do not have low-lying autoionization states
and therefore are not subject to low-temperature dielectronic recombination
(the process discussed in the papers by Nussbamer \& Storey).
Thus, the rates of radiative recombination towards He-like, Li-like and
Na-like ions described here should
be the total recombination rate at nebular temperatures, and only need
to be supplemented by high-temperature dielectronic recombination.

\section{Calculations and fits}
In this paper, we present the calculations and analytic fits
to the rates of
radiative recombination towards H-like, He-like, Li-like and Na-like ions
of all elements from H through Zn ($Z=30$). For calculation of
the ground level recombination rate coefficients for He-like,
Li-like and Na-like ions, we used the updated version of the
analytic fits to the photoionization cross sections (Verner et al. 1995)
which give the corrected threshold cross sections and near-threshold
cross section behavior based on the OP data, and ensure accurate
high energy cross sections by use of the fitting formula with the
correct non-relativistic asymptote. At high temperatures, the direct
recombination to the ground state gives the largest contribution
to the total radiative recombination rate coefficient.

For calculation of the rate
coefficients to the excited levels with $n \le 5$ of the five and
more times ionized species, we used Clark et al. (1986) fits to
the partial photoionization cross sections. Note that the formula
given by Clark et al. (1986) fits the cross sections up to $10E_{\rm th}$
only, where $E_{\rm th}$ is the ionization threshold energy, and does not
ensure correct asymptotic behavior: the cross sections decrease too
slowly with increasing energy. It leads to artificially too high
recombination rate coefficients for Na-like species at $T \sim 10^8$~K.
To avoid it, we substituted Clark et al. high energy tails by
fits $\propto E^{-3}$ from $10E_{\rm th}$ to $100E_{\rm th}$,
and by non-relativistic asymptote $\propto E^{-3.5-l}$, where $l$ is
a subshell orbital quantum number, above
$100E_{\rm th}$. This substitute changes the high-temperature rates
for He-like and Li-like ions by less than 1\% but improves the
rate coefficients for Na-like species at $T \gtrsim 10^8$~K.

For calculation of the radiative recombination rates to the
excited levels with $n \le 10$ of the first four ionization states
where Clark et al. (1986) fits are not valid, we used the
Gould (1978) method of correction for incomplete shielding.
The required values of energy levels with $n \le 10$ for
\ion{He}{1}, \ion{Li}{1}, \ion{Li}{2}, \ion {Be}{2}, \ion {Be}{3},
\ion{B}{3},\ion{B}{4}, \ion{C}{4}, \ion{Na}{1}, \ion{Mg}{2}, \ion{Al}{3},
\ion{Si}{4} were
retrieved from the TOPbase 0.7 (Cunto et al. 1993). We used the
hydrogenic approximation for levels with $n>10$ (the species
listed above) and for levels with $n>5$ (five and more times ionized
species).

To estimate uncertainty of the Gould method, we have made more
detailed calculations of the radiative recombination rates for
\ion{He}{1}. This is motivated by the importance of this process
in measurements of the primordial helium abundance (Reeves 1992).
We used the OP
\ion{He}{1} partial photoionization cross sections (Fernley, Taylor, \&
Seaton 1987) of 40 low-lying excited levels with $n<10$ ($l<3$;
$^1S$, $^1P^{\rm o}$, $^1D$, $^3S$, $^3P^{\rm o}$, and $^3D$ levels)
to obtain recombination rates to these levels.
All levels with $n \ge 10$ were assumed hydrogenic.
Then we compared the obtained total radiative recombination rates
with those calculated by the Gould method. At $T \le 10^2$~K, the discrepancies
between these two methods are less than 1\%, since the main contribution to
the total rate at these temperatures comes from the recombination to
highly excited
hydrogenic levels. At $T>10^2$~K, the discrepancies begin to increase
with increasing temperature, and reach a maximum of 10\% at $T=10^5$~K. At
higher $T$, the discrepancies decrease again, since the main
contribution to the total rate at high temperature comes from the
ground state recombination. Thus, we can estimate that the accuracy
of the Gould method is not worse than 10\%.
The comparison of experimental and
fitted photoionization cross sections of neutrals (He, Li, Na)
reveal that the accuracy of the fits to the ground state photoionization
cross sections is better than 10\% also. Therefore, the absolute accuracy
of the calculated total radiative recombination rates presented here
is not worse than 10\% for neutrals, and must be better for ions.

Careful integration of the Milne relation and use of the
photoionization cross sections which are accurate as near the
thresholds as at high energies ensure the high accuracy of
the total recombination rate coefficients in a wide range
of temperature. We calculated all the rates from 3~K to $10^{10}$~K
using non-relativistic photoionization cross sections. However,
at temperatures
above $10^8$~K, relativistic corrections may be important for
high-$Z$ ions.
As we have mentioned, the largest contribution to the
total recombination at $T>10^8$~K comes from the
recombination to the ground level. Our
fits for the ground level photoionization cross sections
of H and He agree within 2\% with the experimental cross sections
(Veigele 1973) at energies up to 1~MeV.
Thus, the given recombination
rates coefficients are accurate at temperatures
up to $10^{9}$~K for low-$Z$ ions. For high-$Z$ ions, $T=10^8$~K
is more secure upper limit.

We fitted all the total rate coefficients obtained in the temperature
range from  3~K to $10^{10}$~K using the formula
\begin{equation}
\alpha_{\rm r}(T)=a\left[\sqrt{T/T_0}\left(1+\sqrt{T/T_0}\right)^{1-b}
\left(1+\sqrt{T/T_1}\right)^{1+b}\right]^{-1},
\end{equation}
where $a$, $b$, $T_0$ and $T_1$ are the fitting parameters.
Formula (4) ensures correct asymptotic behavior of the
rate coefficients (Ferland et al. 1992) as at low as at
high temperatures: $\alpha_{\rm r}(T)\propto T^{-0.5}$ at
$T \ll T_0 \ll T_1$, and $\alpha_{\rm r}(T)\propto T^{-1.5}$
at $T \gg T_1 \gg T_0$. The average accuracy of fitting is better
than 2\% (maximum error is less than 3\%) for He-like ions
and better than 4\%
(maximum error is less than 6\%) for all Li-like and Na-like species.
The errors of fitting are smaller for highly-ionized species.
Note that although all the fits are accurate in comparison with the
calculated data up to $10^{10}$~K, the upper limits of validity
($10^{8}$~K for high-$Z$ ions, and $10^{9}$~K for low-$Z$ ions)
are determined by non-relativistic character of the calculations.

Because of its importance for cosmological problems,
we present two fits for the \ion{He}{1} rates which are calculated
by more detailed method (see above). The first fit is valid at
$T \le 10^6$~K (the rms average error is 2.5\%, maximum error 5\%).
Note that the \ion{He}{1} high-temperature dielectronic recombination
rate at $T=10^6$~K is two orders of magnitude larger than the radiative
recombination rate at this temperature. The second fit is valid for
all temperatures from $T=3$~K up to
$T=10^{10}$~K (the rms average error is 5\%, maximum error 10\%).
The errors of fitting for all species under study
are not higher than the estimated absolute errors.
Therefore, our fitting does not lead
to any loss of accuracy in comparison with the calculated data.

We also fitted the recombination rate coefficients for all the H-like ions.
Generally, the rate coefficients for all hydrogenic species can be obtained
from the \ion{H}{1} rate coefficient by use of the relation
\begin{equation}
\alpha_{\rm r}(Z,T)=Z\alpha_{\rm r}(1,T/Z^2).
\end{equation}
However, for user's convenience we provide the separate fits to
all the hydrogenic species which ensure the uniform accuracy of
the fitting for all ions in the same temperature range
3~K--$10^{10}$~K. The rms relative errors of the fits for all
hydrogenic species are less than 0.5\%, the maximum errors are less
than 1\%.

The fit parameters are listed in Table 1. We have made extensive
comparisons of our fits with the fits provided by AP,
Shull \& Van Steenberg (1982),
Arnaud \& Rothenflug (1985), Landini \& Monsignori Fossi (1990, 1991),
Pequignot et al. (1991) and Arnaud \& Raymond (1992).
Our fits are in a very good agreement with the fits by
Pequignot et al. (1991) and Arnaud \& Raymond (1992)
in the temperature range where the previous fits are valid.
Our fits are accurate in a much wider temperature range, however,
since our fitting formula has correct low-temperature and
high-temperature asymptotes. The agreement with AP,
Shull \& Van Steenberg (1982),
Arnaud \& Rothenflug (1985) and Landini \& Monsignori Fossi (1990, 1991)
is worse, especially for the cases interpolated along
isoelectronic sequences in earlier papers .
Figures 1--6 illustrate the comparison for helium-like ions \ion{O}{7} and
\ion{Si}{13}, lithium-like ions \ion{C}{4} and \ion{Fe}{24}, and sodium-like
ions \ion{Mg}{2} and \ion{Ca}{10}.
In Table 2,
we compare the rates given by our fits with the rates
provided by Gould (1978) at $T=10^4$~K for the three species which we
have in common. The rates for these species obtained from the previously
available fits are also shown.

For the hydrogenic species,  Arnaud \& Rothenflug (1985) recommended
the formula given by Seaton (1959):
\begin{equation}
\alpha_{\rm r}(Z,T)=5.197\times 10^{-14}Z\lambda^{1/2}[0.4288+
0.5\ln(\lambda)+0.469\lambda^{-1/3}],
\end{equation}
where $\lambda=157890Z^2/T$~(K). However, this formula is not valid
at high ($T>10^6Z^2$) temperatures. The hydrogenic fit given by
Pequignot et al. (1991) also has incorrect high-temperature
asymptote. Figures 7 and 8 compare our hydrogenic data and
fits with the previously available fits.

It can be useful for astrophysical applications to separate
the partial recombination rate to the ground state and the recombination
rate to all the excited states. We have fitted the rates of
radiative recombination
to the ground states and the rates of
radiative recombination to all the excited
states by the same fitting formula (4). The corresponding fit parameters
are available from the authors upon request. The accuracy of fitting of the
partial rate coefficients is generally slightly
lower than that of the total radiative recombination rates coefficients.

\section{Discussion}

Our long-term goal is to predict the complete spectrum emitted by the first 30
elements. Fits to the needed photoionization cross sections are presented
by Verner \& Yakovlev (1995) and Verner et al. (1995). Here
we present  accurate fits to the total radiative recombination
rates for 107 ions of 4 isoelectronic sequences, 23\% of the rate coefficients
needed for the 465 possible ions of the first 30 elements.
This is the complete set of species for which simple radiative recombination
is expected to be the dominant channel at nebular temperatures.

Our temperature range was selected with astrophysical needs in mind.
We briefly mention two cases, at either temperature extreme,
where the results presented here affect published results.
In the low $T$ extreme, charge transfer with He$^+$ is the dominant CO
destruction mechanism in photodissociation
regions (PDRs) for most conditions (Tielens \& Hollenbach 1985, Hollenbach
\& McKee 1989, Ferland, Fabian, \& Johnstone 1994).
The He$^+$ abundance is set by the balance
between creation mechanisms (often ionization by cosmic rays or X-Rays)
and radiative recombination.  Our total rate coefficient for radiative
recombination of He$^+$ to He at 300 K, a representative PDR temperature,
is $4.43 \times 10^{-12}$ cm$^3$s$^{-1}$.
This is 25\% larger than the rate used by Hollenbach
\& McKee (1989).
Tests show that this does indeed result in a substantially
greater CO abundance and cooling.  At the opposite temperature extreme,
heating by photoionization of hydrogenic species of the heavy
elements is the main heating mechanism for a gas near 10$^6$ K.
Our rates are often quite different from the
power-law fits (AP, Shull \& Van Steenberg 1982, Arnaud \& Rothenflug 1985)
extrapolated to
high temperatures, and tests show that the photoelectric
heating and resulting electron temperature tend to be lower, by
as much as 20\%.

\acknowledgements{This work was supported by grants from the National
Science Foundation (AST 93-19034) and NASA (NAGW 3315).
We are grateful to K. T. Korista, M. Arnaud and D. G. Yakovlev for
useful discussions. We acknowledge the use of the TOPbase 0.7 database
installed by A. K. Pradhan at the Ohio State University.}


\begin{figure}
\plotone{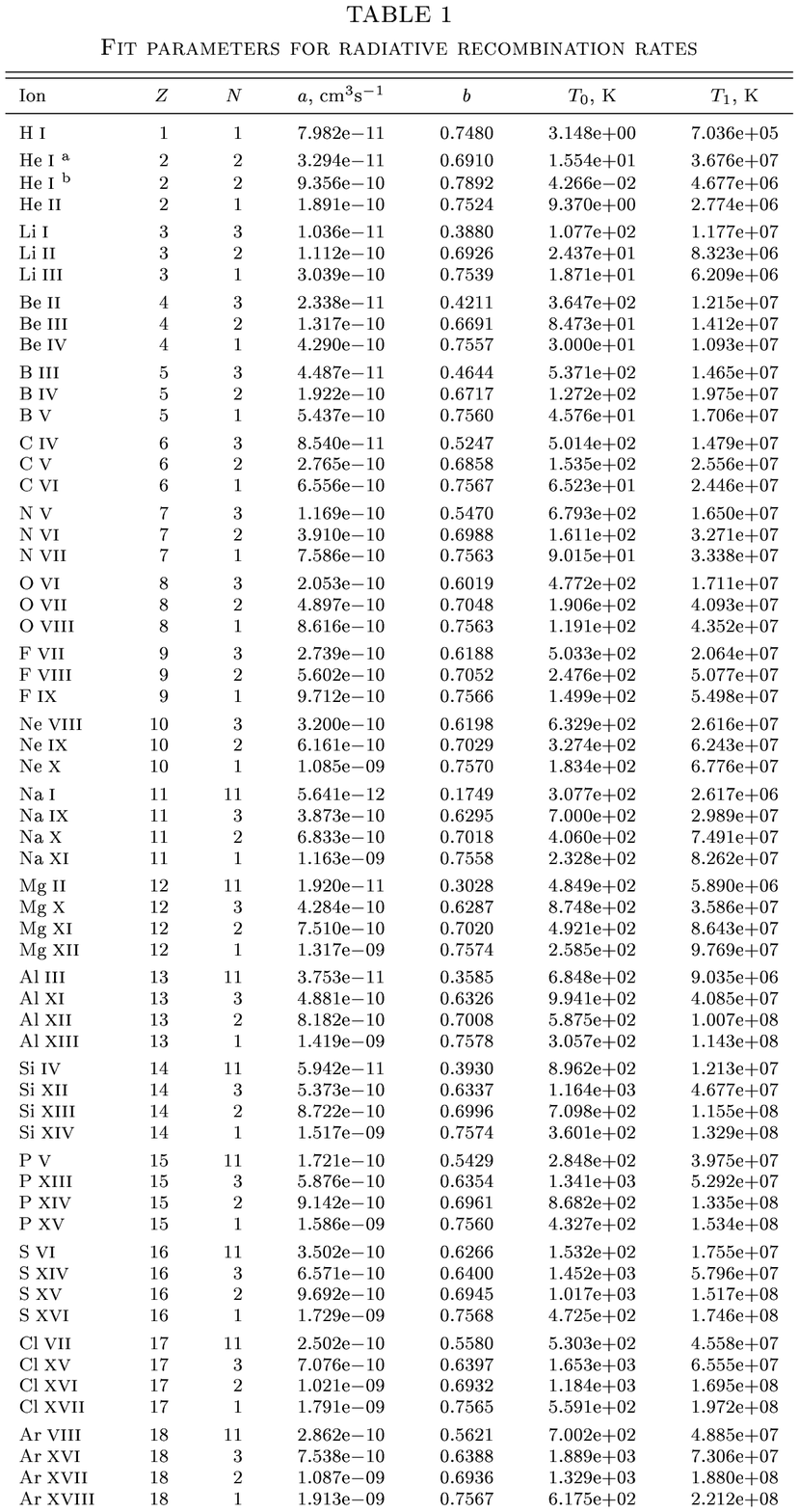}
\end{figure}

\begin{figure}
\plotone{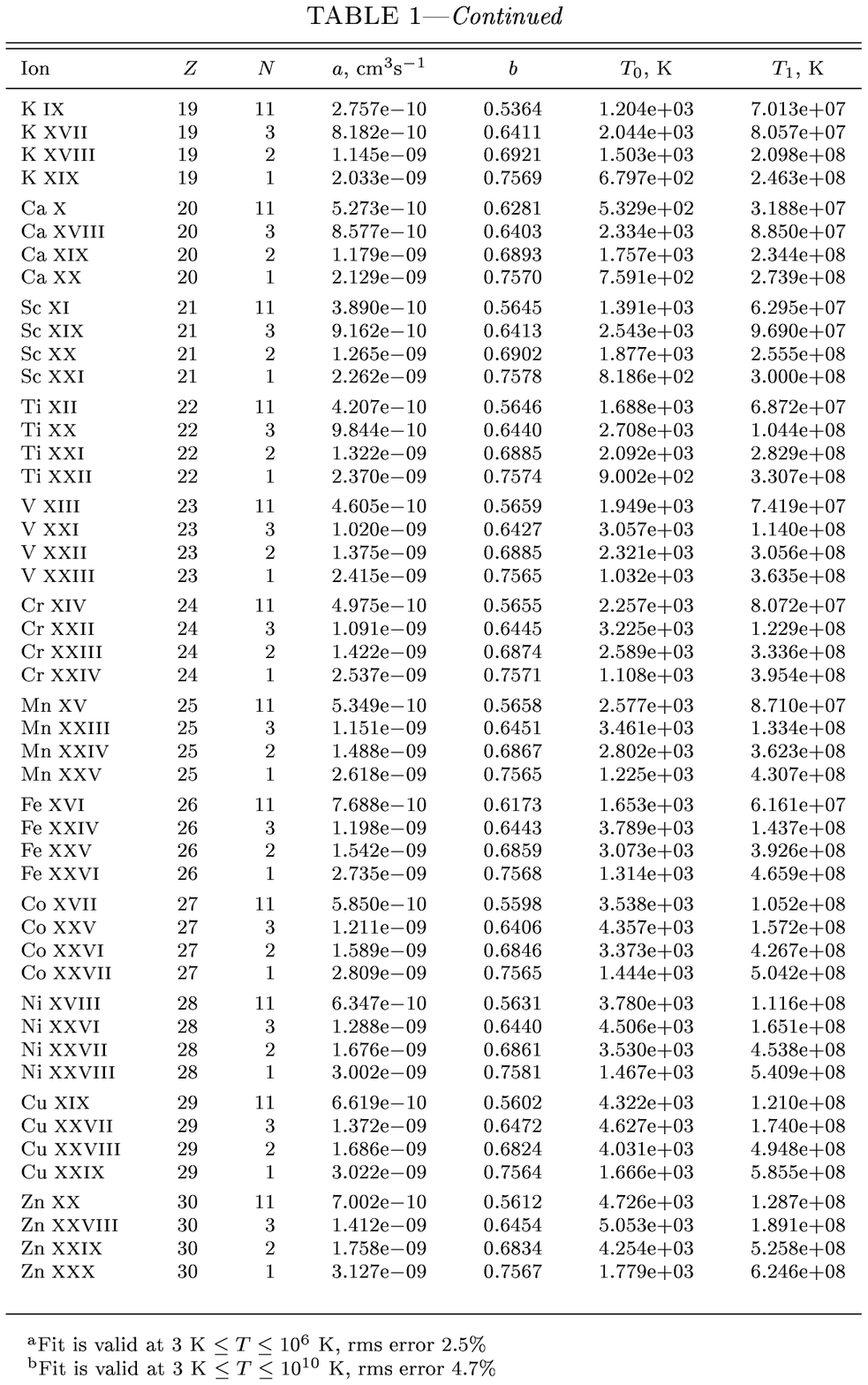}
\end{figure}

\begin{figure}
\plotone{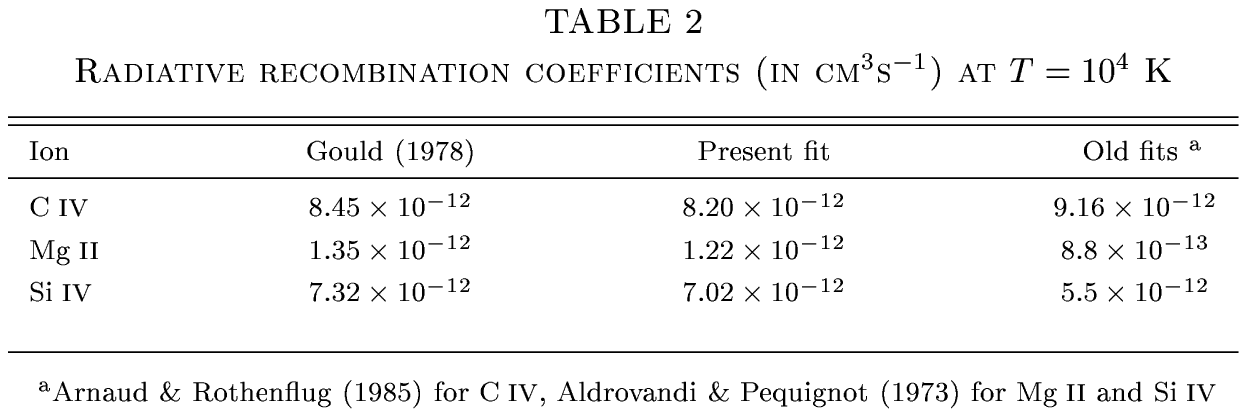}
\end{figure}

\begin{figure}
\plotfiddle{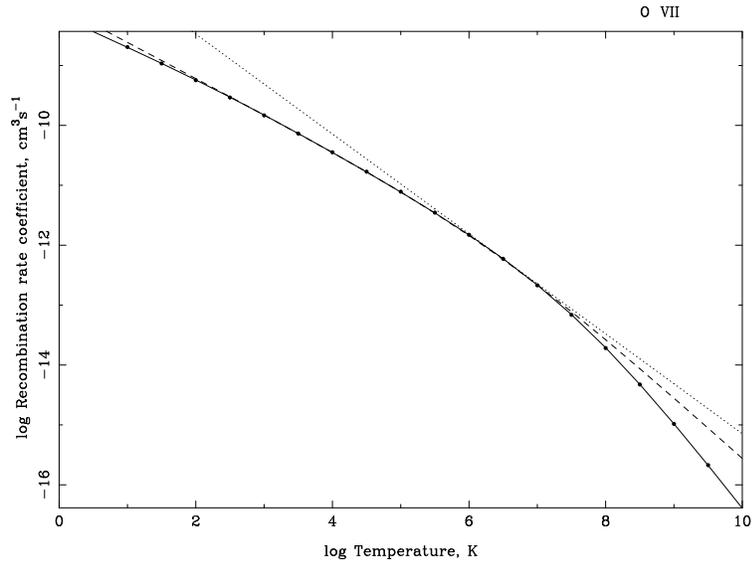}{6.5cm}{-90}{40}{40}{-160}{250}
\caption{Radiative recombination rate coefficient of O VII vs. temperature.
{\it Circles:} present calculations; {\it solid line:} present fit;
{\it dotted line:} Arnaud \& Rothenflug (1985); {\it dashed line:}
Pequignot et al. (1991).}
\end{figure}

\begin{figure}
\plotfiddle{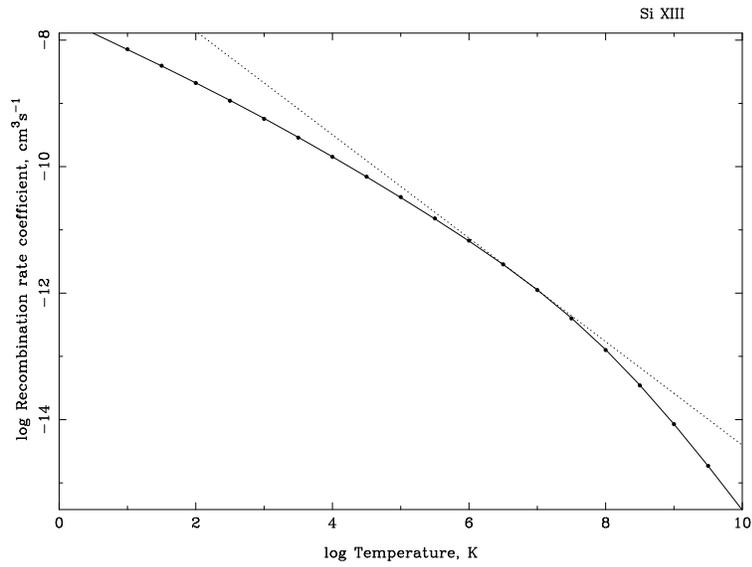}{6.5cm}{-90}{40}{40}{-160}{250}
\caption{Same as Fig. 1 for Si XIII.}
\end{figure}

\begin{figure}
\plotfiddle{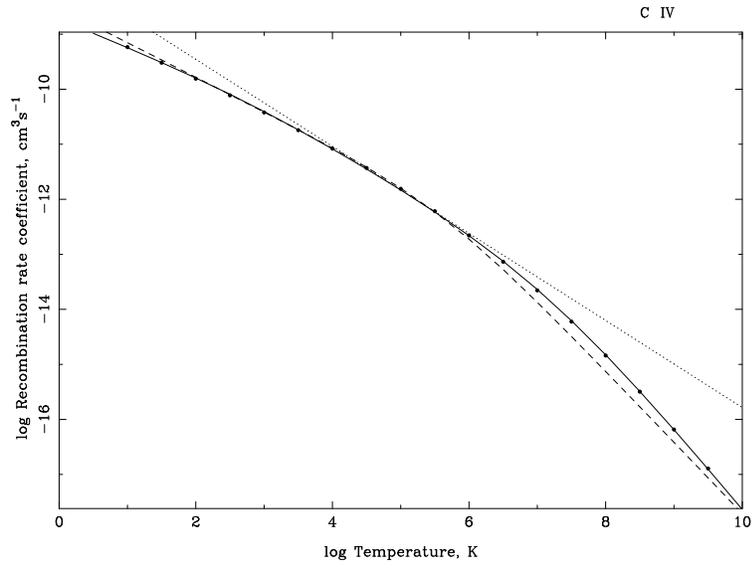}{6.5cm}{-90}{40}{40}{-160}{250}
\caption{Same as Fig. 1 for C IV.}
\end{figure}

\begin{figure}
\plotfiddle{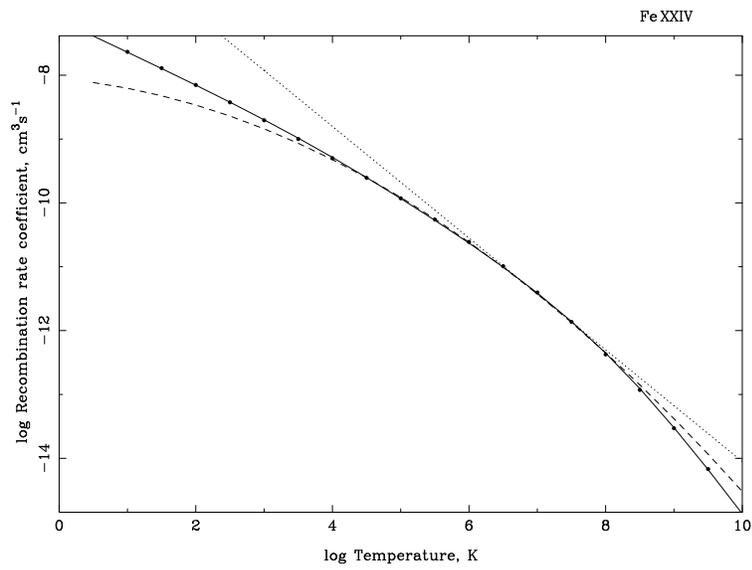}{6.5cm}{-90}{40}{40}{-160}{250}
\caption{Same as Fig. 1 for Fe XXIV. {\it Dashed line:} Arnaud \& Raymond
(1992).}
\end{figure}

\begin{figure}
\plotfiddle{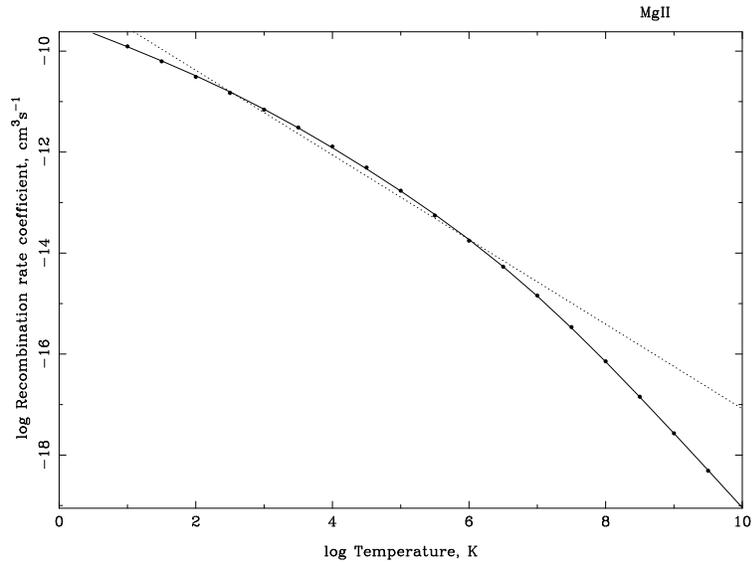}{6.5cm}{-90}{40}{40}{-160}{250}
\caption{Same as Fig. 1 for Mg II. {\it Dotted line:} Aldrovandi \&
Pequignot (1973).}
\end{figure}

\begin{figure}
\plotfiddle{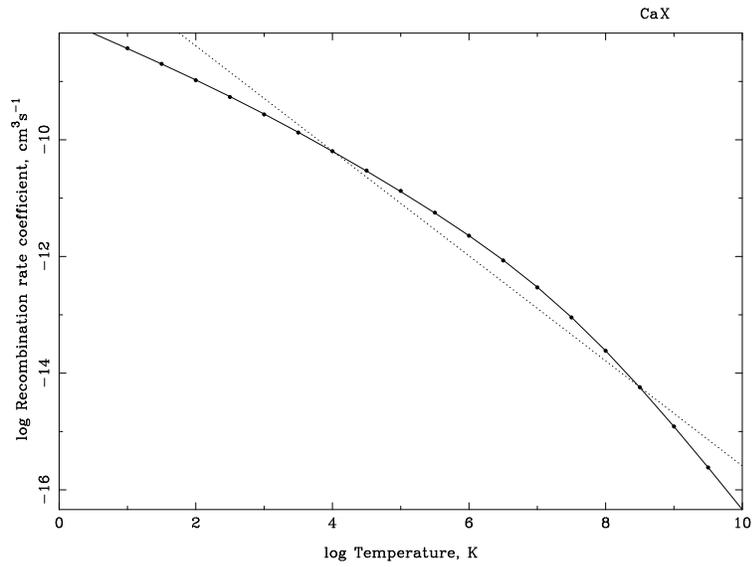}{6.5cm}{-90}{40}{40}{-160}{250}
\caption{Same as Fig. 1 for Ca X. {\it Dotted line:} Shull \& Van Steenberg
(1982).}
\end{figure}

\begin{figure}
\plotfiddle{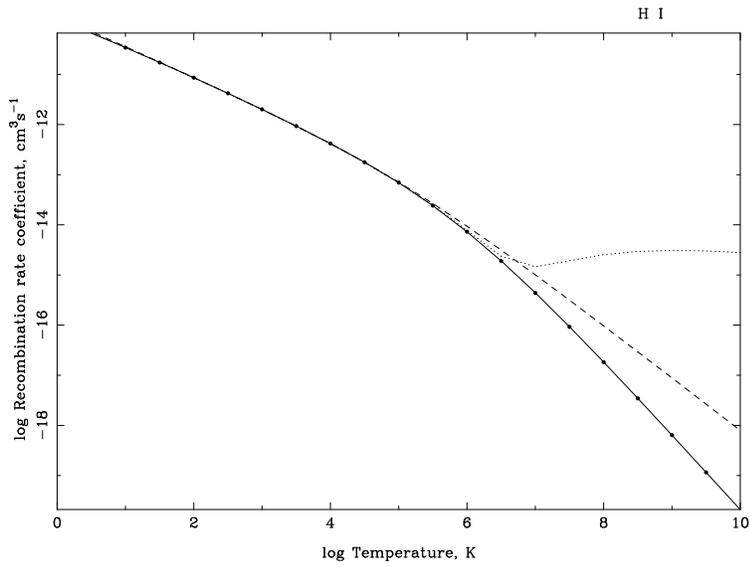}{6.5cm}{-90}{40}{40}{-160}{250}
\caption{Radiative recombination rate coefficient of H I vs. temperature.
{\it Circles:} present calculations; {\it solid line:} present fit;
{\it dotted line:} Seaton (1959); {\it dashed
line:} Pequignot et al. (1991).}
\end{figure}

\begin{figure}
\plotfiddle{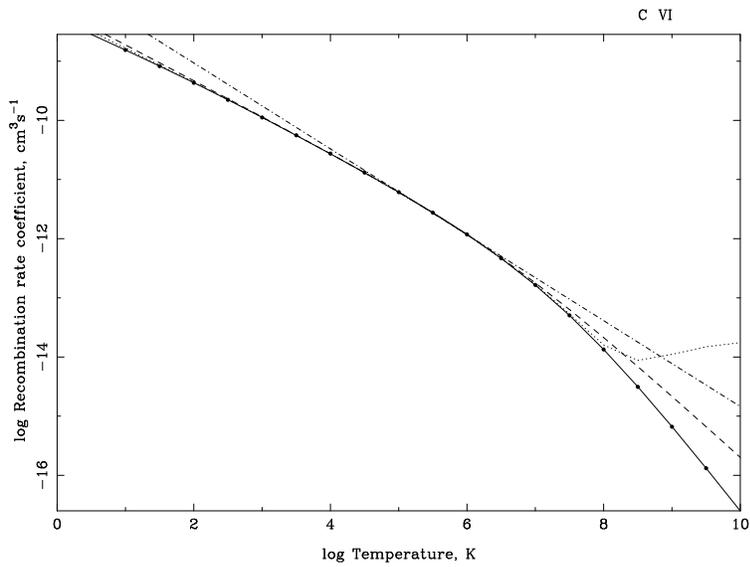}{6.5cm}{-90}{40}{40}{-160}{250}
\caption{Same as Fig. 7 for C VI. {\it Dash-dotted line:}
Shull \& Van Steenberg (1982, see also erratum).}
\end{figure}

\end{document}